\documentclass[paper,11pt]{article}
\usepackage{amsfonts}
\usepackage{amsmath}
\usepackage{amssymb}
\usepackage{graphicx}
\usepackage{lscape}
\usepackage{hyperref}
\usepackage{dcolumn}
\usepackage{bm}
\usepackage{harvard}

\input epsf
\newtheorem{theorem}{Theorem}

\newtheorem{lemma}[theorem]{Lemma}

\begin{document}

\title{ Information-related changes in contact patterns may trigger
oscillations in the endemic prevalence of infectious diseases. }
\author{Alberto d'Onofrio\thanks{%
corresponding author} \\
%EndAName
Division of Epidemiology and Biostatistics,\\
European Institute of Oncology\\
Via Ripamonti 435, 20141 Milano, Italy \\
e-mail: alberto.donofrio@ieo.it, phone:+390257489819 \and Piero Manfredi \\
%EndAName
Dipartimento di Statistica e Matematica Applicata all'Economia,\\
Universit\`{a} di Pisa, \\
Via Ridolfi 10, 5612 Pisa, Italy }
\maketitle

\begin{abstract}
It is well known that behavioral changes in contact patterns may
significantly affect the spread of an epidemic outbreak. Here we
focus on simple endemic models for recurrent epidemics, by modelling the
social contact rate as a function of the available information on the
present and past disease prevalence. We show that social behaviour change
alone may trigger sustained oscillations. This indicates that human behavior might be a critical explaining factor of oscillations in time-series of endemic diseases. Finally, we
briefly show how the inclusion of seasonal variations in contacts may imply
chaos.
\end{abstract}


\section{Introduction}

The prevailing wisdom suggests that besides major factors, such as the
sanitary revolution, or the discovery of vaccines, the role of men in
affecting the disease's dynamics has always been a minor one. This is
mirrored by the amazing simplicity of the role paid to humans in models of
diseases. However, the threat posed by the possibility of a pandemic of
avian flu, and the related need to develop predictive and control tools, 
has clearly indicated that while we dispose of richer and richer models, we
are still poor in the understanding of human behavior~\cite{ferguson1}.
Though serious progresses have been recently made, for instance for the
first time we dispose of a body of standardised international data on social
contact patterns~\cite{mossong}, yet these data give a static picture of the
contacts an individual has in one average day. This fact - human behavior as
static - is postulated in most models of both endemic and epidemic diseases,
but is clearly a coarse abstraction. Individuals are neither static nor
passive: they elaborate the available information and can change their
social behavior to respond to changes in their perceived risk. Symptoms of
behavioral change were evident during the pandemic of Spanish flu ~\cite%
{cummings}. A recent, well documented instance of individual's behavior
change in response to an epidemic outbreak is given by the dramatic decline
in travels and social contacts during the 2003 SARS epidemics in Hong Kong
and Singapore~\cite{ferguson1}. Large behavioral changes were observed in
the S. Francisco homosexual community as a response to the spread of HIV ~%
\cite{mckusick}. Similarly the control of HIV/AIDS in some low resources
settings, for instance the Uganda 'success story' ~\cite{green}, has mainly
been achieved thanks to dramatic changes in sexual beviour, which are now
documented in other Sub-Saharan Africa contexts as well~\cite{gregson}.

The onset of the bioterrorism scare and of SARS have given great impulse to
improve our understanding of the possible effects of the responses by the
public to epidemic threats in the globalisation era ~\cite{fe,cho,de}. The
increased risk of a H5N1 pandemics flu has definitively put better modelling
of human behavior, particularly the individuals' response to an epidemic
threat, at the top of the current research agenda ~\cite{ferguson1}. A
variety of epidemic models including 'thinking agents' who elaborate the
available information, and consequently strategically adapt their social
behaviour not only to changing epidemiological conditions but also to
changes in other agents' behaviour, have been proposed, e.g. \cite{fBauch,Vardavas,fBreban,cummings}. This is a long way from the
first epidemiological model dealing with behavioral changes during an epidemic,
the SIR epidemic model by Capasso and Serio in seventies~\cite%
{capasso,capassobook}. Capasso and Serio allowed the contact rate $\beta$, until then taken as constant, to be a decreasing function of the disease prevalence (i.e. the infective fraction in the total population) $I$. This implies that the Force of Infection (FoI), i.e.  the per-capita rate at which susceptible individuals acquire the infection per unit of time, takes the following non-linear form ~\cite{capassobook}:%
\begin{equation}
FoI(I)=\beta (I)I  \label{capassoFoI}
\end{equation}%
with: $\beta ^{\prime }(I)<0$. Capasso and Serio pointed out that, differently from
standard mass action formulations, this could make the FoI to become a
non-monotone function of the prevalence (e.g. if $\beta (I)=\beta
_{0}(1+hI^{2})^{\text{-}1}$). Capasso and Serio motivated their formulation with
behavioral changes: in epochs of high prevalence the perceived risk of
infection might become very large yielding dramatic changes in individuals'
behavior, therefore also reducing the actual risk of getting the disease.
Since that seminal paper, several other works have been devoted to epidemic
models with a non-linear FoI~\cite{liu,mo,del,li,wa,zxl}, in order to mirror the
existence of some degree of change in contact patterns. However, in some works \cite{liu,li,wa,zxl} (and references therein) it is assumed that contact rate is an increasing function of $I$ in all $\le I \le 1$ \cite{liu,li}, or that it is increasing in an initial interval $0 \le I \le \bar{I} < 1 $ and then decreasing \cite{wa,zxl}. It is worth of note that in these cases multiple equilibria and oscillations, through Hopf bifurcations of endemic states, may arise. We stress, however, that local and global increases of the contact rate (for their biological roots, see  \cite{liu} ) can hardly be related to social phenomena, such as behavioural change, which is the focus of our paper. Indeed it would not be epidemiologically realistic to assume that susceptible subjects do intensify their contacts patterns as the prevalence of the disease increases.\\ Most investigations
of human behavior in relation to diseases dynamics have however been carried
out in relation to STDs, particularly HIV, motivated by the need to
understand the possible impact of information campaigns on sexual behavior,
and as a consequence on infection trends. In particular, the long time
scales of HIV/AIDS have motivated the study of the role of behavioral changes
not only in relation to epidemic control but also under endemic conditions.
Most such papers have incorporated behavioral changes extending in various ways
the phenomenological approach \`{a} la Capasso-Serio~(e.g., among the first papers
appeared on the subject, \cite{velasco}), but there have also been some
works including the individual's behavioral choice within HIV diffusion
models ~\cite{kremer}.

On the other hand no investigation has been carried out on the impact of
behavioral changes on the dynamics of common endemic close-contacts diseases,
such as measles. This is surprising given that, despite the anecdotic
importance paid to pandemics (e.g. of plague), the history, and growth, of
mankind has largely been regulated by devastating recurrent epidemics of
essentially endemic diseases such as smallpox, typhoid fever, or measles. In the
absence of any knowledge on the diseases' etiology and of therapies,
measures of social distancing and behavioral changes were the only walls
against the fatal impact of these diseases. Though most available
documentation regards especially social distancing, as for the plague, there
is evidence that behavioral changes also played a role, as documented in
historical, anthropological and social medicine studies~\cite{fa,hal}.

Now actual forms and extent of such behavioral changes, i.e. the relative
importance of quarantine vs absenteism (to assist sick people) vs reduced
contact rates vs more hygienic practices - are probably largely dependent on
the social, say developed vs developing countries or urban vs rural areas,
and historical context considered. Nonetheless we speculate that,
differently from big but largely spaced epidemics, where knowledge of the
disease impact could only be anecdotic, already in historical times the
adopted mechanisms of behavioral changes for endemic periodic diseases could
have been based on the available information on the state of the disease,
filtered though the knowledge about it, acquired through the social and
family history during past epidemics.

Consistently with this idea in this paper we study a new simple model for
the dynamics of recurrent endemic diseases with behavioral changes, where the
contact rate is a function of some information index $M$ summarising the
current and past history of the disease prevalence:%
\begin{equation}
FoI(M)=\beta (M)I,  \label{ourFoI}
\end{equation}%
where $\beta ^{\prime }(M)<0$, and where $M$ is related to past prevalence
through a suitable function $g(.)$ as follows:%
\begin{equation}
M(t)=\int_{0}^{+\infty }{g(I(t-\tau ))\varrho (\tau )d\tau }.  \label{eqM}
\end{equation}%
where $\varrho (t)$ is a delaying kernel ~\cite{delbook1,delbook2} i.e. a positive function such that: $\int_{0}^{+\infty}\varrho(\tau)d\tau = 1$. The
chosen approach, based on a phenomenological modelling of behaviour change,
is a first step toward more involved modelling of social behaviour change
that has the advantage of providing simpler and clearcut mathematical
results.\newline
A similar approach has been recently used to model a different
behavior-related phenomenon: the social acceptance of vaccination programmes
and rational exemption~\cite{rebaga,n,o,i}. In this case the vaccination
coverage has been represented as an increasing function of $M$.\newline
The fact to relate current contact patterns to the delayed information on
the disease, is realistic also in today's world since for most diseases
information is made available after articulated routine procedures (e.g.
laboratory confirmations), but it was central in historical times due to the
low speed of information spread on the geographic scale.

\section{The model}

Consistently with our aims, we investigate the implications of our
information-dependent FoI on a basic Susceptible-Infectious-Removed (SIR)
model with vital dynamics~\cite{capasso}:%
\begin{eqnarray}
S^{\prime } &=&\mu (1-S)-\beta (M)IS  \label{SIR-M} \\
I^{\prime } &=&\beta (M)IS-(\mu +\nu )I  \label{SIR-S}
\end{eqnarray}%
completed by eq. (\ref{eqM}), governing the dynamics of $M$, and by the
balance equation of the removed fraction $R(t)$: $R(t)=1-S(t)-I(t)$. The
parameters $\mu $ and $\nu $ denote the general mortality rate
and the removal rate from the disease, respectively.\newline
As for $g(I)$, one might take any function such that $g(0)=0$ and $g^{\prime
}(I)>0$. The simplest example is $g(I)=kI$ where $k\in (0,1)$ which could
represent for instance the actually reported incidence of serious case of
the disease.~Another example could be \cite{n,o,i}; $M=wI/(1+qI)
$ where M is a non-linear increasing function of the disease prevalence which
can be taken as a measure of the perceived, rather than actual, risk of
infection ~\cite{rebaga,n,o,i}.\newline
As for the delay kernel $\varrho (\tau )$, some noteworthy cases are the
following ($\delta $ is the Dirac's delta function): \textit{i)} the
no-delay case: $\varrho (\tau )=\delta (\tau )$; \textit{ii)} the fixed
delay: $\varrho (\tau )=\delta (\tau -T)$, with $T>0;$ \textit{iii)} the
Erlang distribution of order $n$:%
\begin{equation}
\varrho (\tau ;a)=\frac{a^{n}}{(n-1)!}\tau ^{n}e^{-a\tau },  \label{Erlang}
\end{equation}%
where the mean delay is given by $T=n/a$ and the standard deviation is $%
\sigma =T/\sqrt{n}$. This model has been largely used in literature ~\cite%
{delbook1,delbook2,aw,gr,ll,kg,ad} since it is more realistic than the case
of constant delay, and it allows a finite dimensional reduction of the
integral equation (\ref{eqM}) to the following $n$ ODEs: 
\begin{eqnarray}
M_{1}^{\prime } &=&a\left( g(I)-M_{1}\right)   \notag \\
M_{j}^{\prime } &=&a\left( M_{j-1}-M_{j}\right) \text{, }j=1,\dots ,n-1, 
  \label{nODEs}
\end{eqnarray}%
where $M=M_{n}$. The case $n=1$ corresponds to the exponentially fading
kernel, the case $n=2$ to the so called strong kernel~\cite%
{delbook1,delbook2}, and the case $n\rightarrow +\infty $ implies $\rho
(\tau ,n/T)\rightarrow \delta (t-T)$.

\subsection{The Basic Reproduction Number}
It is possible to define a basic reproduction number (BRN) ~\cite{capassobook} for our model:%
\begin{equation}
\Re _{0}=\frac{\beta (0)}{\mu +\nu },  \label{brn}
\end{equation}%
which may be interpreted as the average number of secondary cases caused by
a single infectious individual entering a fully susceptible population
that has no information on the disease.\\ In this section we shall show how this parameter influences the behavior of the model in case of generic delay kernel. In fact, it is easy to verify that if $\Re
_{0}\leq 1$ then: 
\begin{equation}
\lim_{t\rightarrow +\infty }\left( S\left( t\right) ,I\left( t\right)
\right) =(1,0),  \label{LAS_DFE_1}
\end{equation}%
i.e. the disease-free state (DFS) is Globally Asymptotically Stable
(GAS). This result immediately follows from the differential inequality: $$%
I^{\prime }\leq I\left( \beta (0)(1-I)-(\mu +\nu )\right). $$ Moreover, since the linearized equation for $I$ is: $$%
i^{\prime }=i\left( \beta (0)-(\mu +\nu )\right), $$ it is easy to show that if $%
\Re _{0}>1$ then the DFS is unstable and it exists a unique endemic state $EE=(S_{e},I_{e})$ where $%
S_{e}=(\mu +\nu )/\beta \left( g\left( I_{e}\right) \right) $ and $I_{e}>0$
is the unique solution of:%
\begin{equation}
\mu \left( 1-\frac{\mu +\nu }{\beta \left( g\left( I\right) \right) }\right)
-(\mu +\nu )I=0.  \label{Ie}
\end{equation}%
Moreover, in case of Erlang kernel, we demonstrate in the appendix that if  $\Re _{0}>1$ then the disease is strongly persistent.\\
Summing up, $\Re _{0}$ indicates that the capability of the disease to invade
the host population or to go extinct is governed by the baseline behavior in
absence of information on the disease spreading, independently on the
structure of the memory kernel.

\section{Oscillations around the Endemic Equilibrium}
As we showed in the previous section, the existence and stability of DFS is
independent of the delay properties, as well as the existence and location
of the endemic equilibrium. On the contrary, the stability properties of
the EE critically depend on $\varrho (\tau )$.\newline
\subsection{Analytical results}
We notice first that in the no-delay case, i.e. when the information index M is related to the current prevalence, $M=g(I)$, if $\Re
_{0}>1$ then the endemic state is GAS, as it was previously obtained by \cite{wa}, although without modelling the information. 
In fact EE is LAS and since: $$div\left(  \frac{S^{\prime }}{I},\frac{I^{\prime }}{I}\right)=-\frac{\mu}{I} - \beta\left(g\left(I\right)\right) + g^{\prime}\left(I\right)\beta^{\prime}\left(g\left(I\right)\right) <0$$
the Dulac's criterion (see the appendix of  \cite{capassobook}) rules out the possibility of closed orbits.
\newline
Second, in the case of the exponentially fading kernel, which leads to the following three-dimensional family of models:
\begin{eqnarray}
S^{\prime } &=&\mu (1-S)-\beta (M)IS  \label{SIRexpS} \\
I^{\prime } &=&\beta (M)IS-(\mu +\nu )I  \label{SIexpI} \\
M^{\prime } &=&a\left( g(I)-M\right)   \label{SIRexpM} 
\end{eqnarray}%
for which some lengthish algebra shows (see appendix) that independently of the specific functional forms
of $g$ and $\beta $ the endemic state is (at least) locally asymptotically
stable (LAS). \\ However, we note that the exponentially fading Kernel gives
maximum weight to the current, i.e. at the time $\tau =0$, rather than past,
disease history. This might be un-realistic in many cases. Therefore, in the
following we consider the Erlang strong kernel $\varrho (\tau )=a\tau
e^{-a\tau }$, so that the integral equation (\ref{eqM}) can be replaced by a
pair of ODE given by(\ref{nODEs}) with $n=2$, leading to the following family of models:
\begin{eqnarray}
S^{\prime } &=&\mu (1-S)-\beta (M)IS  \label{SIRERLS} \\
I^{\prime } &=&\beta (M)IS-(\mu +\nu )I  \label{SIRERLI} \\
M_{1}^{\prime } &=&a\left( g(I)-M_{1}\right)   \label{SIRERLM1} \\
M^{\prime } &=&a\left( M_{1}-M\right)  \label{SIRERLM2}.
\end{eqnarray}%
Thus, in principle, it is possible to analytically characterise the local stability of the endemic state for the above four-dimensional system. However, also for very simple 
$g(I)$ and $\beta (M)$, the problem becomes analytically cumbersome. Some
partial results on the local stability of the endemic state are thus
reported in the appendix for generic $\beta$ and $g$. Though partial these results indicate some
biologically meaningful facts: \textit{i)} for very small delay the endemic
state is locally stable. This makes sense since in this case the system
approaches the undelayed case (therefore the system is also GAS in this
case); \textit{ii)} likewise for large delays the system is LAS;  \textit{%
iii)} the LAS of the endemics state continues to prevail for arbitrary
delays if at equilibrium $\beta (g(.))$ is sufficiently large and $\beta
^{\prime }$ sufficiently small. The latter point means that stability always
prevails if contact patterns at equilibrium are not dramatically altered by
behaviour change as a response to small changes in the endemic prevalence.
\subsection{Numerical results}
\begin{figure}[ptb]
\centering\includegraphics[width=0.95\linewidth,clip]{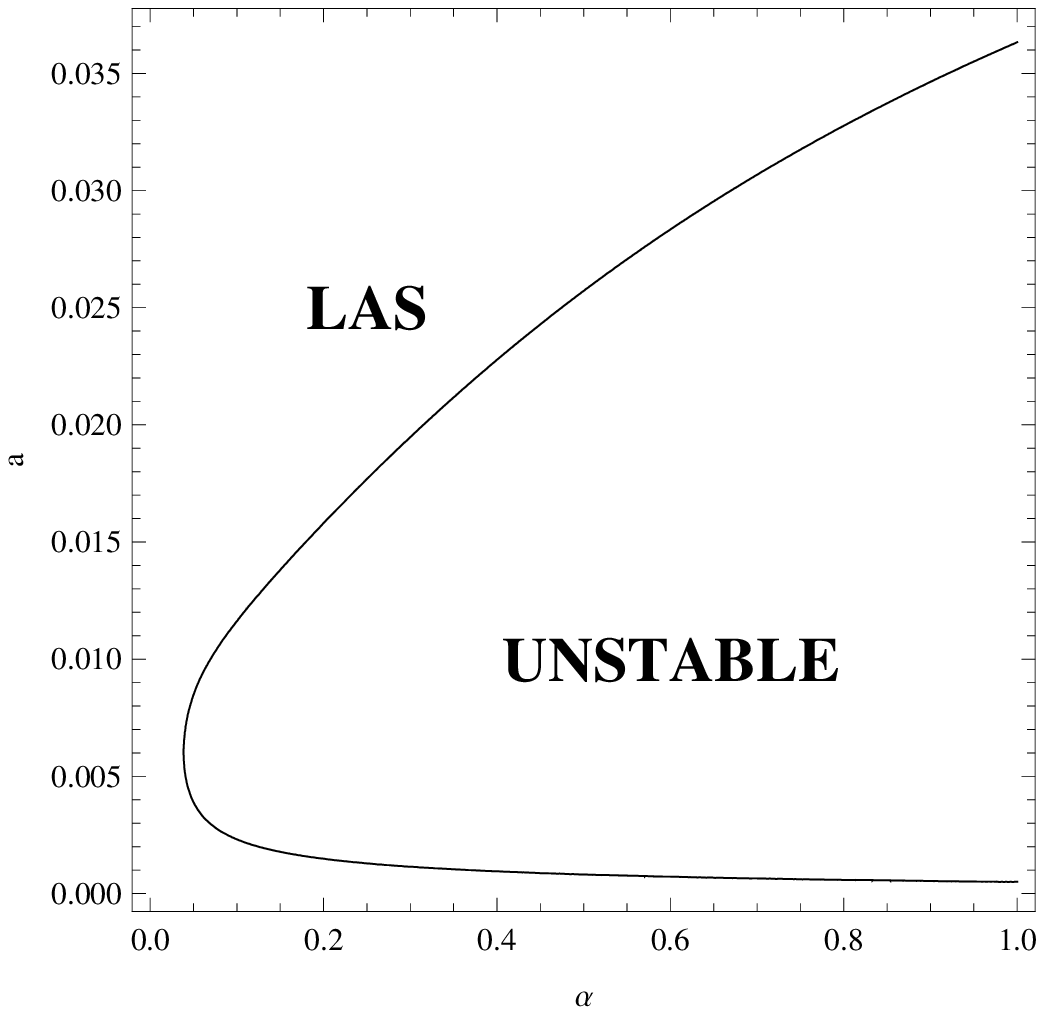} 
\caption{Strong Erlangian kernel. Local stability boundary, and corresponding regions of Local Asymptotic stability and instability in the parameter space  $(\alpha,a)$), where $\alpha$ is the relative rate of decline of the contact rate for an infinitesimal increase of the infective prevalence, and $a$ is inversely related to the average information delay $T$ ($a=2/T$).}
\label{bifu}
\end{figure}

\begin{figure}[ptb]
\centering  \includegraphics[width=0.95\linewidth,clip]{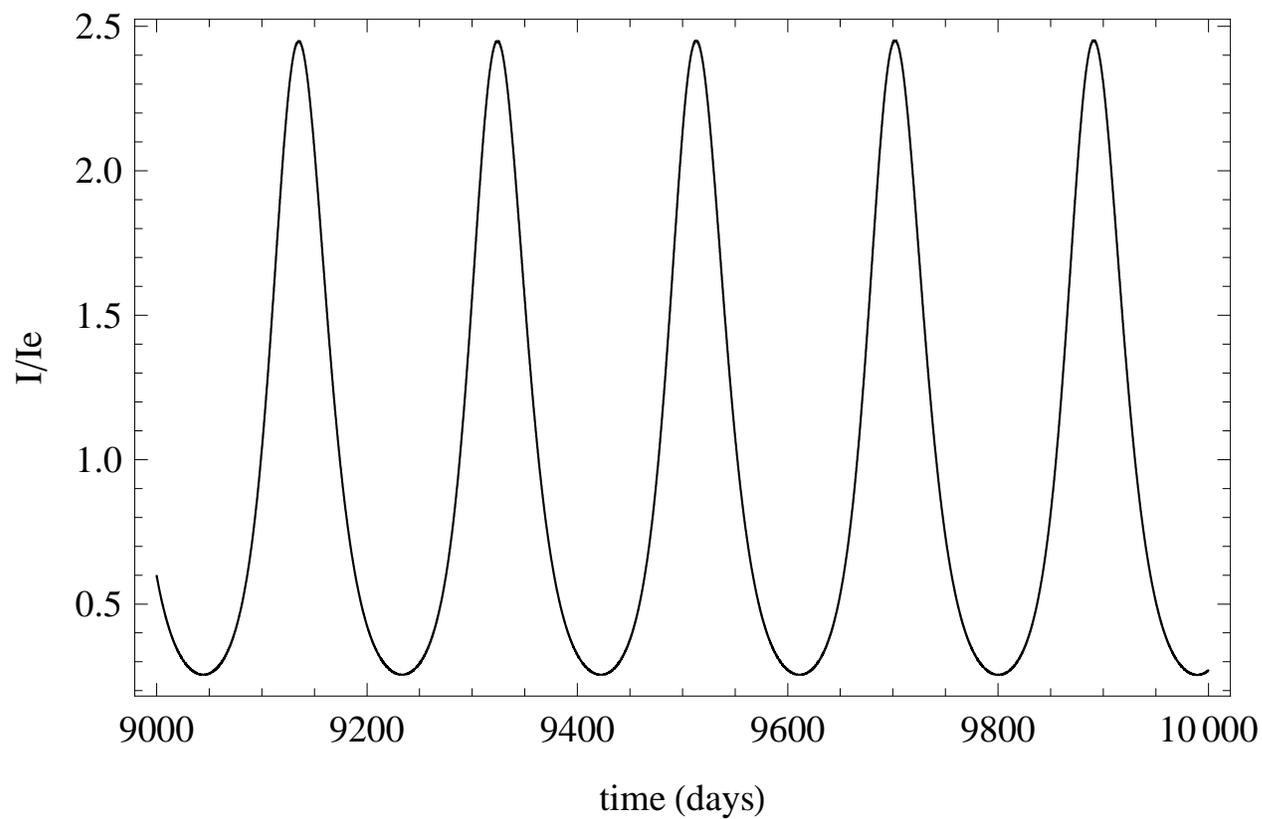} 
\caption{After-transient sustained oscillations of infectious fraction (scaled
to the equilibrium value $I_{e}(\protect\alpha)$) that are induced by
information-dependent FoI. Here $\protect\alpha=1$ and $a=30^{-1}$ days$^{-1}
$ corresponding to an average memory of two months: $T=60$ days}
\label{figata}
\end{figure}

To get further insight in the system behaviour, in particular as regards the
possibility of instability triggered by social behaviour change, we shall
investigate numerically the 'benchmark' case $g(I)=kI$ and%
\begin{equation}
\beta (M)=\frac{\beta _{0}}{1+\alpha \frac{M}{I_{\ast }}},  \label{anstaz}
\end{equation}%
where the parameter $I_{\ast }$ is simply a reference value, which we introduce to keep $\alpha$  small.
This choice of $\beta(M)$ allows to analytically calculate $I_{e}$:%
\begin{equation}
I_{e}(\alpha ,k)=I_{SIR}\frac{I_{\ast }\Re_{0}}{\frac{k\alpha \mu }{\mu +\nu }%
+I_{\ast }\Re_{0}}  \label{I_ee}
\end{equation}%
where $I_{SIR}=\mu(1-(\mu+\nu)/\beta_0)/(\mu+\nu)$ denotes the equilibrium in the classical mass-action SIR
model with contact rate $\beta _{0}$ \cite{capassobook}, and $S_{e}(\alpha ,k)=1-(1+\nu /\mu
)I_{e}(\alpha ,k)$. Note that the presence of behavioral changes affects the
location of the endemic state, compared to the standard SIR model, by
lowering the endemic prevalence $I_{e}$ and increasing the susceptible
fraction $S_{e}$. As bifurcation parameters we choose $\alpha $ and $a$ that
are the new parameters introduced by our model. The former tunes
unambiguously the degree of change of contact patterns: given the level of
the prevalence $I$, the greater $\alpha $ is, the smaller the contact rate
as a function of $I$. In particular: $\alpha $ represents the relative rate
of decline of the contact rate for an infinitesimal increase of the
infective prevalence;  $a$ tunes, as already pointed out, the delaying kernel.\newline
In our simulations, we consider the following measles-like parameter
constellation: $\mu =(75\cdot 365)$days$^{-1}$, (implying a life expectancy $%
A=1/\mu =75$ years), $\nu =1/7$ days$^{-1}$, implying an average duration of
the infectious period of one week, and $k=1$. As for the contact rate, we
set $\beta _{0}=20(\mu +\nu )$, which implies $\Re _{0}=20$. Finally we set $%
I_{\ast }=0.9\mu /(\mu +\nu )$ which is the endemic prevalence in the
classical SIR model with a BRN equal to 10, implying $I_{e}=I_{\ast }$. Note
that at the steady state the value of the contact rate is one half of the
value at $M=0$. This choice of parameters is sufficiently realistic and it
allows a straightforward comparison with the the behavior of the classical
SIR model.\newline
Figure \ref{bifu} reports the the Hopf stability boundary in the $(\alpha ,a)
$ space and the related stability/instability regions. The figure shows that
for low $\alpha $ ($\alpha <\alpha _{min}\approx 0.0386$) the endemic state
is LAS independently of the delay, whereas for $\alpha >\alpha _{min}$ the
equilibrium is unstable for low or high values of the mean delay, and it is
unstable for intermediate delays. For example for $\alpha =0.5$ instability
prevails for an average memory lenght $T=2/a$ between $76$ days and $3.36$
years, whereas for $\alpha =1$ it prevails for $T$ between $55.4$ days and $%
10.4$ years. Finally, at Hopf curve there are supercritical Hopf
bifurcations.\newline
In figure \ref{figata} we show, for $\alpha =1$, the off-transient behavior
of the ratio $I(t)/I_{e}$ for an average memory length of two months ($a=1/30
$ days$^{-1}$). As it is easy to see, there are sustained oscillations such
that the maximum peak is about $2.4$ times the value that would have been
reached in case of instantaneous reaction to the disease prevalence. Quite
interestingly, the minimum value is roughly equal to one quarter of $I_{e}$.
This detail is of some interest, since the values of the minima reported in
literature for regimes of sustained oscillations are usually very small. In
this circumstance low populations stochastic effects should be taken into
account. Furthermore, increasing the average delay both reduces the minimum
of $I(t)$ and increases the period of the oscillations. For example, for $T=6
$ months and $\alpha =0.50$ to $1$, the oscillation becomes biennial. Thus,
seemingly plausible values of the average delay $T$ and of the rate of
decline of the contact rate $\alpha $ have the potential to generate the
classical biennial inter-epidemic period of measles. \newline
Finally, as in the classical SIR model~\cite{ly,arsc,ea}, nonlinear resonance and chaos may be triggered by seasonal variations of the contact
patterns~\cite{grfr,kg,bac,waz}, modelled by one-year periodically modulating the
FoI: 
\begin{equation*}
FoI(M,t)=\varphi (t)\beta (M)I.
\end{equation*}%
For example, by assuming the simple seasonal law: 
\begin{equation*}
\varphi (t)=1+\varphi _{1}cos\left( \omega t\right) 
\end{equation*}%
where $\omega =2\pi /$(1 year)), one has that for $T=6$ months and $\alpha =1$
the behavior of the system for $\varphi _{1}=0.3$ is chaotic and the Maximum
Liapunov Exponent ~\cite{mle} of the associated Poincare's map is equal to $%
3.067$ (value calculated by using the software NDT by Dr. Joshua Reiss \cite%
{jr}). 

\section{Concluding Remarks}
The results of this paper show facts that, at the best of our knowledge,
were never pointed out in the literature on endemic diseases. The main one
regards the issue of oscillations, a central one in theoretical
epidemiology. As well documented, a variety of diseases, first of all
pre-vaccination measles in developed countries, exhibit steady oscillations
in large communities, with a period around two years for the measles.
Nonetheless, the classical SIR (and SEIR) model has a unique GAS endemic
state where oscillations are only a transient, non sustained, phenomenon~%
\cite{capassobook}. A huge work has therefore been devoted to the
explanation of regular and chaotic oscillations in time-series of infectious
diseases ~\cite{kg,ly,arsc,ea,grfr} as the 'output' of external periodic
forcing due to periodic changes in contacts rates, plus further concurring
factors such as time heterogeneity in susceptible recruitment. The present
paper indicates that oscillations of endemic disease could have been
triggered by a further, largely neglected, factor, i.e. changing human
social behavior as a response to the recurrent disease threat. More
precisely it shows that the existence of some degree of behavioral changes in
response to the disease threat, coupled with some information delay as a
memory of the past disease history may be a source of oscillations of common
SIR diseases. Comparing this result to the oscillations triggered by
changing vaccination behavior~\cite{rebaga,n,o,i}, where an exponentially
fading memory is sufficient to destabilize the endemic state and trigger
sustained oscillations, here it happens that memories more focused in the
disease past history, as exemplified by the Erlangian strong kernel, are
needed. Moreover, here sustained oscillations occur even under moderate
values of the parameter $\alpha $ tuning the individuals' reaction to
changing prevalence of the disease. In particular, in the oscillatory regime,
seemingly plausible values of the average delay $T$ and of the rate of
decline of the contact rate $\alpha $ have the potential to generate a large
range of values of the inter-epidemic period of the sustained oscillation,
including annual and biennial oscillations, as commonly observed for
pre-vaccination measles. In the end, although empirical work is needed to
better support our results, all this suggests the interesting possibility
that behavioral changes could have been important factors in shaping endemic
profiles of diseases in history, as recently suggested by ~\cite{ferguson1}.

\section{Aknowledgements} We thank very much two anonymous referees and an anonymous associated editor whose suggestions were very helpful.

\section{Appendix}
\textit{If the referees should think it appropriate, this appendix might be put online.}

\subsection{Persistence of the disease}
As we mentioned in the main text, its hold that:
\begin{lemma} If $\Re_0 >1$ then the disease is strongly persistent, i.e. it exists a $\epsilon_0>0$ such that if $I(0)>0$, $S(0)>0$ and $M(0)>0$  then:
$$ MinLim_{t \rightarrow + \infty}I(t)\ge \epsilon_0>0\textrm{, }MinLim_{t \rightarrow + \infty}S(t)\ge \epsilon_0>0\textrm{, }MinLim_{t \rightarrow + \infty}M(t)\ge \epsilon_0>0 .$$
\end{lemma}
For the sake of the notation simplicity, we proof this lemma in the case of weak Erlang
kernel, but the proof substantially holds for all Erlang delay kernels.\newline We start noticing that the set:
\begin{equation}\label{pi}
\Omega = \left\{(S,I,M)| S\ge 0\textrm{, }I\ge 0\textrm{, }S+I \le 1\textrm{, }0\le M \le g(1)\right\}
\end{equation}
is positively invariant for our model. Moreover, the disease free state $DFS=(1,0,0)$ is on $\partial \Omega$, as well as its stable manifold which is the set $\left\{(S,I,M)\in \Omega | I=0\right\}$. Thus, in virtue of the Freedman-Ruan-Tang theorem \cite{frt}, if $DFS$ is unstable (i.e. if $\Re_0>1$) and, of course, the initial point is in the interior of $\Omega$, then the state variables are strongly persistent.
\subsection{LAS of the EE under the exponentially fading memory}
In case of exponentially fading kernel, linearizing the system at the
endemic equilibrium $(S_{e},I_{e},g(I_{e}))$, the characteristic polynomial
is: 
\begin{equation*}
\lambda ^{3}+(a+\mu +I_{E}\beta (g(I_{E})))\lambda ^{2}+
\end{equation*}%
\begin{equation*}
+\lambda \left( I_{E}(\mu +\nu )\beta (g(I_{E}))+a\left( \mu +I_{E}\beta
(g(I_{E}))-\frac{aI_{E}(\mu +\nu )g^{\prime }(I_{E})\beta ^{\prime
}(g(I_{E}))}{\beta (g(I_{E}))}\right) \right) +
\end{equation*}%
\begin{equation*}
a\left( I_{E}(\mu +\nu )\beta (g(I_{E}))-\frac{I_{E}\mu (\mu +\nu )g^{\prime
}(I_{E})\beta ^{\prime }(g(I_{E}))}{\beta (g(I_{E}))}\right) ,
\end{equation*}%
whose coefficients are all positive.\\
Thus the Routh-Hurwitz condition for the local stability of the endemic
state is: 
\begin{equation*}
\left( \mu +I_{E}\beta (g(I_{E}))-\frac{I_{E}(\mu +\nu )g^{\prime
}(I_{E})\beta ^{\prime }(g(I_{E}))}{\beta (g(I_{E}))}\right) a^{2}+\left(
(\mu +I_{E}\beta (g(I_{E})))^{2}-I_{E}^{2}(\mu +\nu )g^{\prime }(I_{E})\beta
^{\prime }(g(I_{E}))\right) a+
\end{equation*}%
\begin{equation*}
+I_{E}^{2}(\mu +\nu )\beta (g(I_{E}))^{2}+I_{E}\mu (\mu +\nu )\beta
(g(I_{E}))>0,
\end{equation*}%
which is fulfilled in all cases since the coefficients of the above second order
polynomial in $a$ are positive for all positive increasing $g(I)$ and
positive decreasing $\beta (M)$. As a consequence, EE is locally
asymptotically stable.

\subsection{LAS of the EE under the strong Erlangian kernel}

In this case one may show that the Routh-Hurwitz condition takes the form: 
\begin{equation*}
RH= \sum_{h=1}^{5}q_{h}a^{h}>0
\end{equation*}%
where the rhs is a 5-th order polynomial in $a$ with null zero degree coefficient and: 
\begin{equation*}
q_{1}=2I_E^{2}(\mu +\nu )^{2}\beta (g(I_E))^{2}(\mu +I_E\beta (g(I_E)))>0
\end{equation*}%
\begin{equation*}
q_{5}=2\mu +2I_E\beta (g(I_E))-\frac{2I_E(\mu +\nu )g^{\prime }(I_E)\beta ^{\prime
}(g(I_E))}{\beta (g(I_E))}>0
\end{equation*}%
since $\beta ^{\prime }(M)<0$ and $g^{\prime }(I)>0$. The positivity of $%
q_{1}$ implies that for sufficiently small $a$ (i.e. for large average
delays) the EE is LAS. The positivity of $q_{5}$ implies, on the contrary,
that for sufficiently large $a$ (i.e. for very small average delays) the EE
is LAS. Moreover, for large $a$ it also holds that:
\begin{equation*}
a^{-1}M_{1}^{\prime }\approx 0\Rightarrow M_{1}=g(I)
\end{equation*}%
\begin{equation*}
a^{-1}M^{\prime }\approx 0\Rightarrow M=M_{1}\Rightarrow M=g(I),
\end{equation*}%
Therefore by singular perturbation approximation the system reduces
to the undelayed one, which is not only LAS but also GAS. As far as the
other coefficients are concerned, they have variable sign: 
\begin{equation*}
q_{2}=4I_E(\mu +\nu )\beta (g(I_E))(\mu +I_E\beta (g(I_E)))^{2}+\frac{I_E(\mu +\nu
)\left( \mu ^{2}-I_E\nu \beta (g(I_E))\right) g^{\prime }(I_E)\beta ^{\prime
}(g(I_E))(\mu +I_E\beta (g(I_E)))}{\beta (g(I_E))}
\end{equation*}%
\begin{equation*}
q_{3}=2(\mu +I_E\beta (g(I_E)))\left( \mu ^{2}+I_E\beta (g(I_E))(4\mu +2\nu +I_E\beta
(g(I_E)))\right) +
\end{equation*}%
$$ +\frac{2I_E(\mu +\nu )\left( \mu ^{2}+I_E\beta (g(I_E))(\mu +\nu
-I_E\beta (g(I_E)))\right) g^{\prime }(I_E)\beta ^{\prime }(g(I_E))}{\beta (g(I_E))} $$
\begin{equation*}
q_{4}=4(\mu +I_E\beta (g(I_E)))^{2}-\frac{I_E^{2}(\mu +\nu )^{2}g^{\prime
}(I_E)^{2}\beta ^{\prime }(g(I_E))^{2}}{\beta (g(I_E))^{2}}+\frac{I_E(\mu +\nu )(\mu
-3I_E\beta (g(I_E)))g^{\prime }(I_E)\beta ^{\prime }(g(I_E))}{\beta (g(I_E))}
\end{equation*}%
However, a sufficient condition to have $q_{2}>0$ is: 
\begin{equation}
\mu ^{2}-I_E\nu \beta (g(I_E))<0  \label{cq2p}
\end{equation}%
Similarly we have the following a sufficient condition guaranteeing that $q_{3}>0$: 
\begin{equation}
\mu ^{2}+I_E\beta (g(I_E))(\mu +\nu -I_E\beta (g(I_E)))<0  \label{cq3p}
\end{equation}%
and $q_{4}>0$: 
\begin{equation}
\frac{\beta (g(I_E))\left( \mu -3I_E\beta (g(I_E))-\sqrt{17\mu ^{2}+I_E\beta
(g(I_E))(26\mu +25I_E\beta (g(I_E)))}\right) }{2I_E(\mu +\nu )g^{\prime }(I_E)}<\beta
^{\prime }(g(I_E))<0.  \label{cq4p}
\end{equation}%
As a consequence, if at the EE all the conditions (\ref{cq2p}), (\ref{cq3p})
and (\ref{cq4p}) are met then the EE is LAS. These conditions are quite
general, since they are not linked a priori to particular forms of the
contact rate or of $g(I)$. Although, they are complex and of difficult
reading, a simple biological interpretation is given in the text.

\end{document}